# IT ambidexterity driven patient agility and hospital patient service performance: a variance approach

Rogier van de Wetering[1*]

[1] Department of Information Sciences, Open University of the Netherlands, 6419 AT Heerlen, The Netherlands

*Correspondence: rogier.vandewetering@ou.nl

# IT ambidexterity driven patient agility and hospital patient service performance: a variance approach

## Abstract

Hospitals are currently exploring digital options to transform their clinical procedures and their overall engagement with patients. This paper investigates how hospital departments can leverage the ability of firms to simultaneously explore new IT resources and practices (IT exploration) as well as exploit their current IT resources and practices (IT exploitation), i.e., IT ambidexterity, to adequately sense and respond to patients' needs and demands, i.e., patient agility. This study embraces the dynamic capability view and develops a research model, and tests it accordingly using cross-sectional data from 90 clinical hospital departments from the Netherlands through an online survey. The model's hypothesized relationships are tested using Partial Least Squares (PLS) structural equation modeling (SEM). The outcomes demonstrate the significance of IT ambidexterity in developing patient agility, positively influencing patient service performance. The study outcomes support the theorized model can the outcomes shed light on how to transform clinical practice and drive patient agility.

*Keywords*

## Introduction

Health information technology (HIT) plays a crucial role in the daily medical practice of hospitals. Technologies like the electronic medical record (EMR), decision-support systems, big data analytics, the Internet of Things, and social media apps are only a handful of innovative technologies currently changing and shaping hospitals' healthcare practices. Many hospitals currently embrace the 'digital transformation' as they explore the best suitable digital options and transform their clinical procedures and their overall engagement with patients [1-3]. Hospitals embrace patient-centeredness in the modern-day and age while trying to anticipate turbulent market conditions and regulatory pressures. In this process, hospitals leverage innovative HIT to enhance efficiencies, deliver high quality of care by effectively deploying their HIT assets, resources, and organizational capabilities, and focus on the state-of-the-art patient service delivery [4-6]. However, limited research focuses on leveraging HIT to enhance patient satisfaction and services and drive the departments' performance [7-9]. Therefore, understanding the facets that drive HIT investment benefits in clinical practice is very valuable. Moreover, limited attention has been given to HIT's role in developing specific organizational capabilities to respond to patient's needs and wishes adequately, i.e., patient agility and enhance patient engagement [10-12]. Thus, substantial gaps remain in the extant literature.

The current paper, therefore, addresses two critical limitations. *First*, this current paper unfolds how hospital departments—that are responsible for patient care delivery— can develop the ability to balance 'exploration' and 'exploitation' in IT resource management, i.e., IT ambidexterity [13], to drive a hospitals' patient agility, conceptualized as a dynamic capability. Gaining these insights is important as there seems to be less consensus about IT resources' pivotal role in developing these dynamic capabilities that offer organizations the ability to deliver business services distinctively and mobile to anticipate market disruptions and business changes [14-16]. *Second*, this study shows the impact of patient agility on the department's patient service performance. Focusing on ambidexterity and agility benefits on the department levels is crucial as it has been seldom explored [13, 17, 18], and the literature predominately focused on the organizational level. This study, therefore, foresees that IT ambidexterity will enable the hospital department's ability to adequately 'sense' and 'respond' to patient needs, demands, and opportunities within a turbulent and fast-paced hospital ecosystem context [15, 16, 19]. Gaining these insights is essential, as hospitals are currently very active in exploring their digital options, innovations and transforming their clinical processes and their interactions with patients using digital technologies [2].¬

Against this background and the literature's current gaps, this paper's main objective is to examine if IT ambidexterity contributes to higher patient agility levels and the department's patient service performance. Hence, this research attempts to address the following research questions:

*(1) "To what extent does IT ambidexterity affect the hospital departments' patient agility and, thus, its ability to sense and respond timely and adequately to the patient's needs and demands? Furthermore,*
*(2) what is the role of patient agility in converting the contributions of IT ambidexterity to the department's patient service performance?*

## Theoretical background and research model

### The concept of IT ambidexterity

Ambidexterity refers to the simultaneous alignment of 'exploration' and 'exploitation' by organizations that provide them with sustained competitive benefits [20]. Within information systems research, IT ambidexterity can be conceived as "*..the ability of firms to simultaneously explore new IT resources and practices (IT exploration) as well as exploit their current IT resources and practices (IT exploitation)*" Lee et al. [13]. Hence, IT exploration concerns the organization's efforts to pursue new knowledge and IT resources [13, 21]. IT exploitation captures the extent to which organizations exploit existing IT resources and assets, e.g., reusing existing IT applications and services for new patient services and reusing existing IT skills [13, 22]. IT ambidexterity is considered a key strategic priority and gained serious attention over the past few years. In practice, the simultaneous alignment of IT resources is crucial in forming digital-driven capabilities [1, 4, 23], even so in healthcare [24]. However, the imbalance between exploration and exploitation can lead to suboptimal business results [25]. Therefore, organizations need to adapt existing IT resources to the current business environment and demands and focus on developing IT resources that contribute to long-term organizational benefits [11, 19, 21].

### Dynamic capabilities view and patient agility

The DCV is considered by many scholars to be a leading theoretical framework and is built from a multiplicity of theoretical roots [26]. Dynamic capabilities can be considered "*the organizational and strategic routines by which firms achieve new resource configurations as markets emerge, collide, split, evolve, and die*" [27]. Within the DCT framework, organizations seek a balance between strategies to remain stable in delivering current business services distinctively and mobile to anticipate and effectively address market disruptions and business changes [15, 28]. The DCV, thus, regards the environment as a crucial element that needs to be considered while deploying the firm's strategy. Dynamic capabilities allow firms to remain stable in delivering current business services distinctively and mobile so that they can anticipate and effectively address market disruptions and business changes [15, 29]. These dynamic capabilities have been defined and conceptualized as sets of measurable and identifiable routines that have been widely validated through empirical studies [28, 30, 31].

Organizational agility, or a 'sense-and-respond' capability, has been defined and conceptualized in many ways and through various theoretical lenses in the IS literature [32, 33]. It is also conceived as a manifested type of dynamic capability [15]. The concept is influential among agility studies published in the management and information systems journals, see, for instance, [13, 32]. Organizational agility can be conceived as a dynamic capability if "*they permit organizations to repurpose or reposition their resources as conditions shift*" [34]. Organizational agility allows organizations to respond to changing conditions while proactively enacting the dynamic environment regarding customer demands, supply chains, new technologies, governmental regulations, and competition [15]. The extant literature has conceptualized organizational agility as a higher-order construct [13, 16, 19]. Two organizational routines can be synthesized: 'sensing' and 'responding' to business events in capturing business and market opportunities, synthesizing from the extant literature. Hence, this article perceives patient agility as a higher-order manifested type of dynamic capability that allows hospital departments to adequately 'sense' and 'respond' to patient needs, demands, and opportunities within a turbulent and fast-paced hospital ecosystem context [15, 16].

### Hypothesis development

IT can facilitate hospitals' capability-building and gaining IT business value in the current turbulent market [19, 32, 35]. However, IT business value does not result from the isolated deployment of (non)IT resources and competencies. Instead, it seems to emerge from the complementarity to assimilate and re-align the IT resource portfolio to the changing business needs and demands [36]. Therefore, hospital departments must embrace an ambidextrous IT implementation strategy so that short-term exploitation of (existing) IT resources is balanced with an exploratory mode that drives IT-driven business transformation [37].

*IT exploration* is explicitly about experimenting with and using new IT resources (e.g., clinical decision-support systems, big data analytics, and clinical analytics, Internet of Things, and social media) to serve as a basis to reshape processes and overall patient engagement. IT exploration can help identify and obtain digital technologies and critical IT skills that contribute to the department's strategic ambitions and plan. Also, IT exploration facilitates hospital departments using new digital technologies to adapt and adjust to changing patients' needs adequately and wishes [13, 23]. *IT exploitation*, on the other hand, focuses more on the deliberate enhancement of current IT resources. For instance, think about reusing or redesigning the current EMR for new patient service development and ensuring hospital-wide accessibility to the clinical patient and medical imaging data and information [1, 23]. Furthermore, it enables departments to reuse existing modular and compatible IT-infrastructures and software components and integrate them with their daily business operations and clinical practices [22, 38]. Thus, IT exploration offers hospital departments the ability to make deliberate decisions, enhance the departments' sensing and responding capabilities, and co-evolve with the rapidly changing healthcare market [38]. However, only when the seemingly opposing modes of IT exploration and exploitation, and thus the trade-off approach [21], are in sync, hospital departments are better equipped to improve agility [20, 38]. Based on the above, this study foresees that IT ambidexterity will enable the hospital department's ability to adequately 'sense' and 'respond' to patient needs, demands, and opportunities within a turbulent and fast-paced hospital ecosystem context [15, 16, 19] and defines the following:

> ***Hypothesis 1:*** *IT ambidexterity will positively enhance the patient agility of the hospital department*

Hospitals need to deal with many strategic, organizational, and social challenges, and it has been well understood that focusing on increasing patient service performance is crucial to obtain competitive value and realize the hospitals' ambitions and strategies [3, 39]. Hospital departments can create service value for their patients by leveraging their ability to use their key resources and organizational capabilities [40]. It is essential to comprehend needs, preferences, and wishes to provide patients with compelling propositions and services Anderson, Narus [41]). This reasoning line is also advocated by [42, 43]. Hospital department managers and decision-makers better adopt the patient value perspective that directs the subsequent resource and sense and respond capability, i.e., patient agility, deployment to achieve high service performance levels. By making specific investments in capabilities valued by patients, hospital departments can achieve high levels of patient service performance and value in the turbulent healthcare environment [44].

Effective IT-driven patient agility provides ways for clinicians to improve clinical communication, remotely monitor patients, and improve clinical decision-support [6, 45] and hence improve the patient treatment process and ultimately the quality of medical services [45, 46]. As a result, hospital departments with strong patient agility are more likely to provide service flexibility, high-quality and timely services, achieve patient satisfaction, and improve the accessibility of medical services [16, 23, 47]. Thus, patient agility enables departments to enhance their patient service performance level and ultimately strengthen its market performance [48-50]. Hence, this study hypothesis that:

> ***Hypothesis 2:*** *Patient agility will positively enhance the hospital department's patient service performance.*

Figure 1 shows the focal constructs and the associated hypotheses.

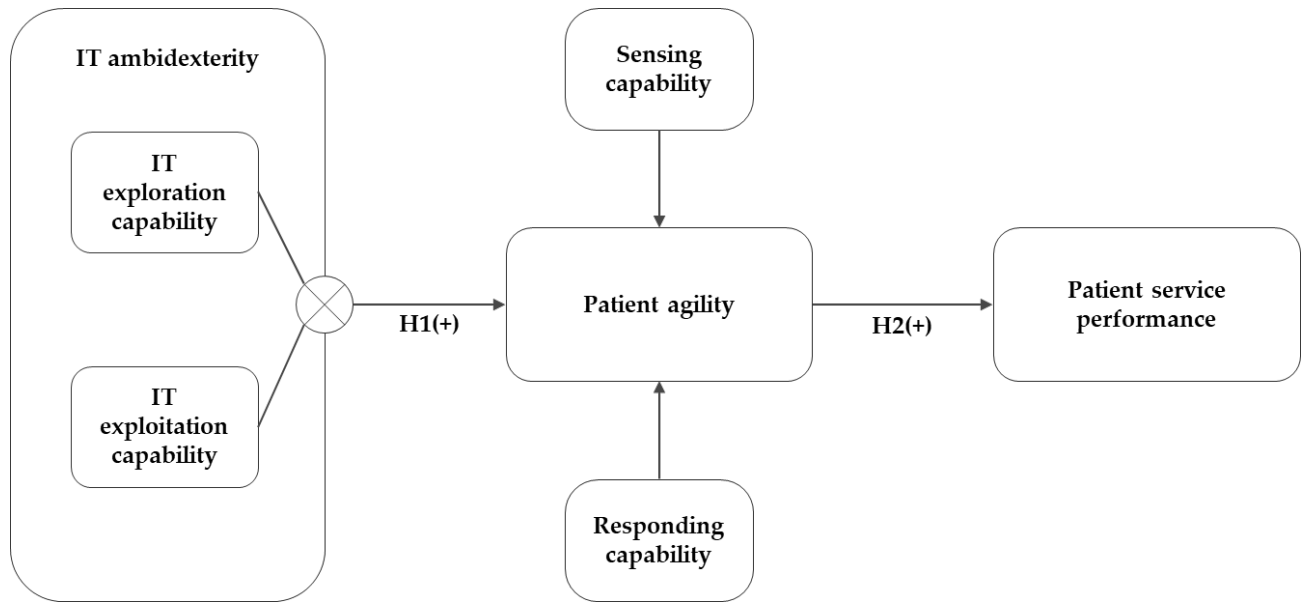

**Figure 1. Research model**

## Methods

### Data collection

Survey data were systematically collected using an online survey that contained all questions to test the study's model and hypothesized relationships. The survey was pretested on multiple occasions by five Master students and six medical practitioners and scholars to improve both the content and face validity of the survey items. The medical practitioners all had sufficient knowledge and experience to assess the survey items effectively to provide valuable improvement suggestions. The data were finally cross-sectionally collected during a field study. The target population was (clinical) department heads- and managers, team-leads, and doctors under the assumption that, at the hospital department level, these respondents are actively involved in contact with patients or at least have an intelligible insight into the department's patient interactions, and the use of IT. Data were conveniently sampled from Dutch hospitals through the Master students' professional networks within Dutch hospitals. The data we collected between November 10$^{th}$, 2019, to January 5$^{th}$, 2020. Also, anonymity for the respondents was guaranteed. This study uses 90 complete survey responses for final analyses. Within the obtained sample, 28.9% of the respondents work for a University medical center, 41,1% work for a specialized top clinical (training) hospital, and 30% work for general hospitals. Most survey respondents are medical heads/chairs of the department (51.1%)[1], 24.4% is a practicing doctor, 11.1% is department manager, while the remaining 13.3% hold other positions such as specialized oncology nurses.

This research targets single respondents and is, therefore, sensitive to possible biases. Possible method bias was accounted for following specific guidelines by Podsakoff et al. [51]. Hence, this study accounted for possible non-response bias using a t-test (between early respondents and the late respondents) to assess whether there is a significant difference in the Likert scale questions' responses. No significant could be detected. Finally, Harman's single-factor analysis was applied using exploratory factor analysis (in using IBM SPSS Statistics™ v24) to restrain possible common method bias [51]. Hence, the current study sample is not affected by method biases, as no single factor is attributed to most of the variance.

### Partial least squares approach

This study relied on a variance-based approach to structural equation modeling (SEM). The research model is assessed using Partial Least Squares (PLS) SEM application, i.e., SmartPLS version 3.2.9. [52] for both reliability and validity of the constructs (i.e., 'measurement model') and the hypothesized relationships' assessment as part of the 'structural model' evaluation [53]. PLS is a commonly preferred method when research models include mediation and when the study's nature is exploratory and emphasizes prediction-oriented work [54]. Following Hair et al. [54], we checked if the current sample's suitability using *G*\*Power for power analyses [55], although PLS is typically recommended with relatively small sample sizes. As input parameters, this study assumed a standard 80% statistical power and a 5% probability of error, while the maximum number of predictors in the research model is two. Based on these parameters, a sample of at least 34 cases was needed (a priori) to detect an explained variance (i.e., $R^2$) of at least 0.21. Hence, the obtained sample is sufficient to assess the study's research mode and obtain stable PLS outcomes.

---

[1] 5 respondents claimed that they were team leads.

*Measurements*

This study tried to use empirically validated measures where possible. Also, this study includes only measures that were suitable for departmental-level analyses. *IT ambidexterity* is operationalized using the item-level interaction terms of IT exploration (ITEXPLORE) and IT exploitation (ITEXPLOIT) [13, 20]. Items were adopted from [13]. *Patient agility* was conceptualized as a higher-order dynamic capability comprising the dimensions '*patient sensing capability*' and '*patient responding capability*' [16, 19, 32]. This study adopts five measures for each of these two capabilities based on Roberts and Grover [16]. This study builds upon the concept of IT-business value creation [40, 56-58] to conceptualize *patient service performance* (PSP). Thus, consistent with balanced evaluation perspectives, patient service performance is represented by three measures, i.e., enhanced quality, improved accessibility of medical services, and achieving patient satisfaction [35, 40, 59].

A typical seven-point Likert scale (ranging from 1: strongly disagree to 7: strongly agree) was used for each item. Following prior IS and management studies, we controlled for 'size' (full-time employees), operationalizing this measure using the natural log (i.e., log-normally distributed) and 'age' of the department (5-point Likert scale 1: 0–5years; 5: 25+ years). Table 1 includes all the constructs' items.

| Construct | | Measurement item | λ | μ | Std. | *Reliability statistics* |
|---|---|---|---|---|---|---|
| *ITEXPLORE* | | *Please indicate the ability of your department to: (1. Strongly disagree–7. Strongly agree* | | | | |
| | EXLR1 | Acquire new IT resources (e.g., potential IT applications, critical IT skills) | 0.86 | 4.01 | 1.67 | CA: 0.79 CR:0.86 AVE:0.60 |
| | EXLR2 | Experiment with new IT resources | 0.92 | 3.81 | 1.62 | |
| | EXLR3 | Experiment with new IT management practices | 0.89 | 3.43 | 1.62 | |
| *ITEXPLOIT* | EXPT1 | Reuse existing IT components, such as hardware and network resources | 0.91 | 5.29 | 1.28 | CA:0.85 CR:0.90 AVE:0.68 |
| | EXPT2 | Reuse existing IT applications and services | 0.94 | 5.18 | 1.32 | |
| | EXPT3 | Reuse existing IT skills | 0.95 | 5.13 | 1.25 | |
| *Sensing* | | *Indicate the degree to which you agree or disagree with the following statements about whether the department can (1 – strongly disagree 7 – strongly agree)* | | | | |
| | S1 | We continuously discover additional needs of our patients of which they are unaware | 0.89 | 4.10 | 1.66 | CA:0.89 CR:0.92 AVE:0.71 |
| | S2 | We extrapolate key trends for insights on what patients will need in the future | 0.77 | 4.43 | 1.63 | |
| | S3 | We continuously anticipate our patients' needs even before they are aware of them | 0.89 | 4.03 | 1.68 | |
| | S4 | We attempt to develop new ways of looking at patients and their needs | 0.79 | 4.72 | 1.52 | |
| | S5 | We sense our patient's needs even before they are aware of them | 0.86 | 3.94 | 1.66 | |
| *Responding* | R1 | We respond rapidly if something important happens with regard to our patients | 0.93 | 4.52 | 1.50 | CA:0.91 CR:0.93 AVE:0.89 |
| | R2 | We quickly implement our planned activities with regard to patients | 0.91 | 4.52 | 1.42 | |
| | R3 | We quickly react to fundamental changes with regard to our patients | 0.92 | 4.54 | 1.53 | |
| | R4 | When we identify a new patient need, we are quick to respond to it | 0.87 | 4.11 | 1.62 | |
| | R5 | We are fast to respond to changes in our patient's health service needs | 0.87 | 4.76 | 1.71 | |
| | | *We perform much better during the last 2 or 3 years than comparable departments from other hospitals in: (1. Strongly disagree–7. Strongly agree).* | | | | |
| PSP | PSV1 | Achieving patient satisfaction | 0.83 | 4.98 | 1.32 | CA:0.75<br>CR:0.85<br>AVE:0.66 |
| | PSV2 | Providing high-quality service | 0.85 | 5.28 | 1.25 | |
| | PSV3 | Improving the accessibility of medical services | 0.75 | 4.80 | 1.33 | |

**Table 1. Construct, items, and reliability statistics**

# Results

## *Measurement model analyses using PLS*

Three types of tests were done to assess the SEM model's measurement model through SmartPLS v 3.3.3. [52], i.e., (1) internal consistency reliability, (2) convergent validity, and finally (3) discriminant validity test [53, 54].

Cronbach's alpha and the composite reliability estimation show that all values are above the 0.7 thresholds, demonstrating sufficient reliability [53]. Also, this study assessed construct-to-item loadings. None of the items had to be removed as all loadings were above 0.70 [60][2]. The average variance extracted (AVE) values were used to assess convergent validity. The threshold for acceptable values is 0.50 [61]. The obtained AVE values from SmartPLS all exceed this threshold. Finally, discriminant validity was assessed through three well-known but different tests. In the first step, cross-loadings were investigated. High cross-loading, i.e., correlations of items (related to one specific latent construct) on other constructs can negatively impact discriminant validity [62]. Outcomes show that all items load substantially more strongly on their intended constructs. Assessment of the Fornell-Larcker criterion was used in a second step. In this process, the square root of the AVEs of all constructs is compared with cross-correlation. This analysis shows that the square root values are all higher than the correlation with other latent constructs [54]. Additional evidence for discriminant validity was found in a final step by subjecting the data to heterotrait-monotrait (HTMT) metric analysis [63]. Results show acceptable HTMT outcomes far below a conservative 0.90 upper bound. The higher-order (formative) construct of patient agility was assessed using variance inflation factors (VIFs) values for the constructs patient sensing and patient responding capability. These VIF-values were well the conservative threshold of 3.5. Hence, no multicollinearity is present within the research model [64].

## *Hypotheses testing*

This study used a non-parametric bootstrap resampling procedure [53] to test the hypotheses. Hence, support was found for the first hypothesis, i.e., IT ambidexterity positively impacts patient agility ($\beta = 0.48$; $t = 6.48$; $p < 0.0001$). Also, support was found for the second hypothesis, i.e., patient agility → patient service performance ($\beta = 0.47$, $t = 6.11$, $p < 0.0001$). Specific mediation guidelines [51] were followed to investigate the model's imposed mediation effects. Outcomes show that patient agility 'fully' mediates the effect of IT ambidexterity on patient service performance [53, 65]. Also, control variables shows non-significant effects: 'size' ($\beta = -0.01$, $t = 0.01$ $p = 0.92$), 'age' ($\beta = -0.01$, $t = 0.14$, $p = 0.89$).

The explained variance for patient agility is 23% ($R^2 = .23$) and 22% of the variance for patient service performance ($R^2 = .22$). A subsequent blindfolding assessment for the endogenous latent constructs using Stone-Geisser values ($Q^2$) shows that the model has predictive power [53]. The $Q^2$ values far exceed 0, i.e., patient agility ($Q^2 = 0.21$) and patient service performance ($Q^2 = 0.14$). Figure 2 summarizes the final structural model results.

---

[2] Only one measurement item in the survey had a loading of 0.68; this is still in the range of acceptable item loadings.

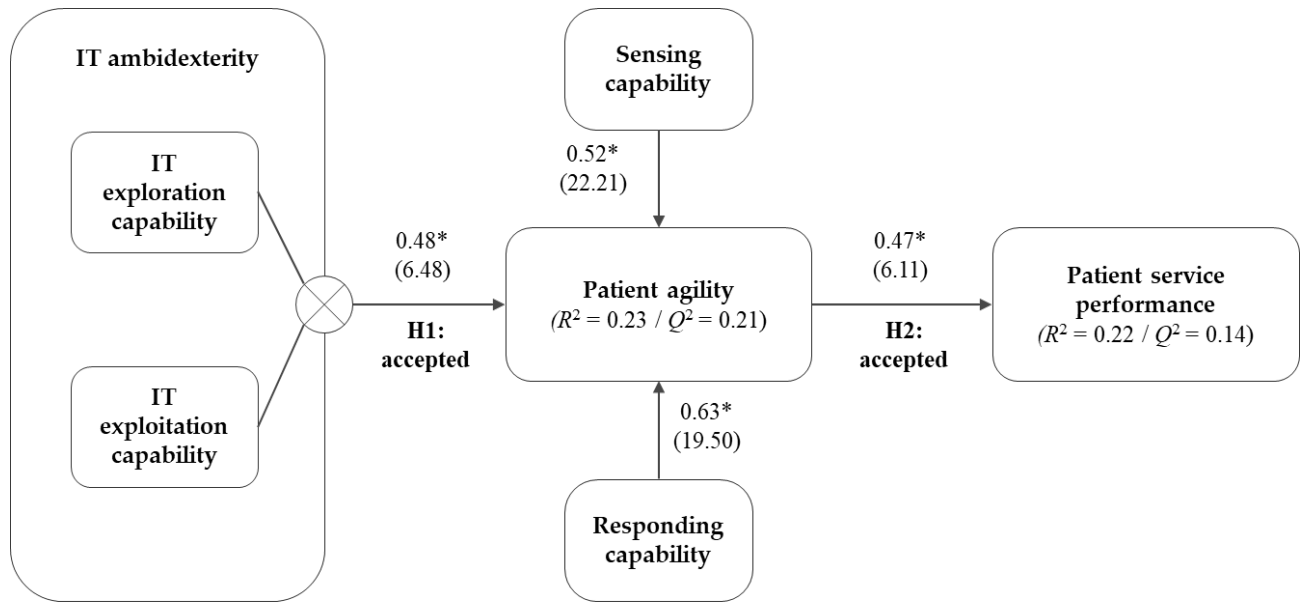

**Figure 2. Structural model results**

## Discussion and concluding remarks

Modern hospitals need to ensure that their processes can meet the needs of an increasingly complex environment, especially now during the COVID-19 crisis. Under these acute conditions, it is essential to maintain strategic flexibility so that adequate digital options and sensing and responding behavior are exercised [19, 66, 67]. However, it is well known that many hospitals currently struggle in their digital transformation efforts, and this process is usually painfully slow, with many hurdles to overcome. Therefore, this study tried to unfold how IT ambidexterity enhances the hospital department's patient service performance through patient agility. This study makes several contributions to theory. First, this study's central claim was that when hospital departments are ambidextrous in IT resources management, they are more likely to sense and respond adequately to patients' needs and wishes and achieve better patient service performances. The results of this study corroborate this claim based on the PLS analyses. Therefore, the results shed light on the current lacunas in the extant literature concerning the mechanisms through which patient service performance and benefits can be achieved through IT ambidexterity and patient agility. Second, this study adds to the current knowledge base on how 'digitizing' supports the capability-building processes, facilitates patient agility, and contributes to the much-needed insights on obtaining value from IT at the departmental level [13, 17, 18, 68]. Third, we extend the study by Lee et al. [13] by showing that IT ambidexterity directly positively affects the department's organizational ability to sense and respond.

This study offers several implications for practice. First, this study shows that hospital departments should invest in their capability to balance both the organization's efforts to pursue new knowledge and IT resources and their capability to take advantage of existing IT resources and assets. Empirical results show that IT ambidextrous hospital departments can better identify, develop new innovative digital opportunities and patient services, and enhance patient agility. Thus this study unfolds the critical resources and capabilities that hospital departments can leverage from a patient agility perspective. Results suggest that hospitals that are committed to the process of ambidextrously managing their IT resources are more proficient in promptly sensing and responding to patients' medical needs and demands. Therefore, decision-makers can justify HIT investments as a source of IT's business value [13, 57, 69]. Second, many hospital departments become highly practiced and thus stuck at achieving the benefits and patient service outcomes they achieve today. When the department's doctors and decision-makers want different or better patient service outcomes, it is essential first to diagnose the current interplay of IT ambidexterity, patient agility capabilities, i.e., sensing and responding in the department, to identify key drivers and barriers to the desired change. In practice, attempts to change a department's working way by changing just one steering mechanism nearly always fail. It seems that reinforcing the nature of the current system overwhelms any single change. Effective enhancement initiatives focus simultaneously on changing individuals' behavior and changing institutional features. Therefore, hospital department decision-makers should pay attention to end-users psychological meaningfulness, stakeholder involvement, and providing adequate resourcing and infrastructures when implementing new digital technologies [70-72]. Several study limitations should be addressed. First, the current data were collected using a single informant strategy. Therefore, method bias could be a concern. Future work could use a matched-pair approach where different respondents address independent (explanatory) and dependent constructs. Also, the current sample, although sufficiently for the current study purposes, is relatively small. Hence, a more extensive sample could contribute to the robustness of the results. To conclude, this study provides critical insights into the hypothesized relationship between ambidexterity in IT resource management, i.e., balancing between aiming for exploration (long-term perspective) and exploitation (current business environment perspective) mechanisms by which patient agility can be achieved in practice.

## Acknowledgments

We want to thank Josja Willems, Reinier Dickhout, Rick Smulders, Yves-Sean Mahamit, and Renaldo Kalicharan for their valuable contributions to the data collection and for sharing their perspectives in numerous discussions. Also, we would like to express our gratitude to all participating hospitals. Your active role made this a success.